\documentclass[twocolumn,pre,nofootinbib,showpacs,superscriptaddress,amsmath,amssymb]{revtex4-1}

\usepackage{rotating,graphicx}
\usepackage{subfigure}
\usepackage{enumitem}
\usepackage{color}
\usepackage{dcolumn}
\usepackage{bm}
\usepackage{multirow}
\usepackage[breaklinks=true]{hyperref}
\usepackage{natbib}
\usepackage{bbm}
\usepackage{soul}
\usepackage{breakcites}
\usepackage{relsize}

\definecolor{orange}{rgb}{1,0.5,0}
\definecolor{purple}{rgb}{0.50196078431,0,0.50196078431}
\definecolor{pink}{rgb}{1,0,1}
\definecolor{gray}{rgb}{0.74509803921,0.74509803921,0.74509803921}
\definecolor{darkgreen}{rgb}{0.45, 0.68, 0.2}
\definecolor{cyan}{rgb}{0.1254,0.6980,0.6666}

\newcommand{\be}{\begin{equation}}
\newcommand{\ee}{\end{equation}}
\newcommand{\bea}{\begin{eqnarray}}
\newcommand{\eea}{\end{eqnarray}}

\DeclareMathOperator{\spec}{spec}

\begin{document}

\title{Discovering and quantifying nontrivial fixed points in multi-field models}

\author{A.~Eichhorn}
\email{a.eichhorn@imperial.ac.uk}
\affiliation{
Blackett Laboratory, Imperial College, London SW7 2AZ, United Kingdom}
\author{T.~Helfer}
\email{now at King's College, London.}
\affiliation{
Blackett Laboratory, Imperial College, London SW7 2AZ, United Kingdom}
\author{D.~Mesterh\'azy}
\email{mesterh@itp.unibe.ch}
\affiliation{
  Albert Einstein Center for Fundamental Physics, Institute for Theoretical Physics, University of Bern, Sidlerstrasse 5, 3012 Bern, Switzerland
}
\author{M.\,M.~Scherer}
\email{scherer@thphys.uni-heidelberg.de}
\affiliation{
 Institut f\"ur Theoretische Physik, Universit\"at Heidelberg, Philosophenweg 16, 69120 Heidelberg, Germany
}

\date{\today}

\begin{abstract}
We use the functional renormalization group and the $\epsilon$-expansion concertedly to explore multicritical universality classes for coupled $\bigoplus_i O(N_i)$ vector-field models in three Euclidean dimensions. 
Exploiting the complementary strengths of these two methods we show how to make progress in theories with large numbers of interactions, and a large number of possible symmetry-breaking patterns. 
For the three- and four-field models we find a new fixed point that arises from the mutual interaction between different field sectors, and we establish the absence of infrared-stable fixed point solutions for the regime of small $N_i$.
Moreover, we explore these systems as toy models for theories that are both asymptotically safe and infrared complete. In particular, we show that these models exhibit complete renormalization group trajectories that begin and end at nontrivial fixed points.
\end{abstract}


\maketitle


\section{Introduction}
\label{Sec:Introduction}

The $O(N)$ Wilson-Fisher fixed point appears in a large variety of systems where it controls the universal critical behavior in the infrared (IR) scaling regime \cite{LeGuillou:1977ju,Guida:1998bx,Wilson:1973jj,Wegner:1976bk,Pelissetto:2000ek}.
Generalizations of this universality class appear in the context of coupled-field models, e.g., for the $O(N_1) \oplus O(N_2)$ two-field model \cite{Fisher:1974zz,Kosterlitz:1976zza,Aharony:2002,Aharony:2002zz,Calabrese:2002bm,Folk:2008mi}. Depending on the number of field components $N_i$ and dimension $d$, one finds that different fixed points (FP) govern the IR behavior of the model. Two of these -- the decoupled (DFP) and isotropic fixed point (IFP) -- can be deduced from the existence of the Wilson-Fisher fixed point. While the DFP is characterized by a complete decoupling of the fields and therefore can effectively be regarded as a model for two independent vector fields, the IFP displays a complete symmetry enhancement to an $O(N_1 + N_2)$ rotational symmetry.
However, the two-field model features another so-called biconical fixed point (BFP), that emerges due to the nontrivial interactions between the two field sectors.
The BFP is fully coupled, \textit{i.e.}, mixed interactions are nonvanishing, but it does not show an enhanced symmetry, as the IFP does, see, e.g., Refs.\ \cite{Calabrese:2002bm,Folk:2008mi,Eichhorn:2013zza}.
Interestingly, it turns out that the $O(N_1)\oplus O(N_2)$ model in $d = 3$ dimensions exhibits exactly one IR-stable fixed point (with no more than \textit{two} relevant directions) for any pair of values $N_1$ and $N_2$.
In this work, we address the question whether generalizations of the two-field model to the case of three and four fields allow for further unprecedented fixed-point solutions, that are relevant for the IR scaling behavior of the respective model. 
Moreover, we look for additional confirmation of our previous study of $n = 3$ fields in three dimensions \cite{PhysRevE.90.052129}, where no stable fixed point (with no more than \textit{three} relevant directions) was found for small values of $N_i$ -- in contrast to the two-field model.\footnote{Here, we consider only IR-stable fixed points for which the effective potential is real and satisfies a stability criterion \cite{Eichhorn:2013zza,PhysRevE.90.052129}.}
In our previous study of this system \cite{PhysRevE.90.052129}, we searched for fixed points using the nonperturbative functional renormalization group (RG) \cite{Wetterich:1992yh,Berges:2000ew,Polonyi:2001se,Pawlowski:2005xe,Gies:2006wv,Delamotte:2007pf}. 
Within this scheme, the $\beta$-functions are non-polynomial functions of the couplings and it is therefore challenging to make sure that numerical fixed-point searches do indeed uncover all stable fixed points of the system. 
To address this problem, we match the solutions of the renormalization group $\beta$-functions derived within the framework of the functional RG to those obtained with the Wilsonian momentum-shell RG by employing an expansion in $\epsilon = 4-d$.
The $\epsilon$-expansion features $\beta$-functions that are polynomials of the couplings. A comprehensive study of all fixed points, that are continuously connected to the Gaussian FP at $d = 4$, is therefore straightforward. On the other hand, the full nonlinear $\beta$-functions of the functional RG yield reasonable estimates of the stability of fixed points already at low orders in the approximation.
Indeed, comparing fixed points in both RG schemes, we can convincingly identify stable fixed points and determine their stability regions in the space spanned by the values of the field components $N_i$ in arbitrary dimensions.

More recently, interacting fixed points have become an active field of research in four-dimensional models, which are explored in the context of an ultraviolet (UV) completion for gravity \cite{Weinberg:1980gg,Niedermaier:2006wt} as well as QFTs including matter fields \cite{Litim:2014uca, Litim:2015iea,Esbensen:2015cjw}. In this setting, an interacting fixed point provides a well-defined microscopic starting point from which a fundamental quantum field theory (QFT) valid on all scales, can be defined. Here, we add another example to the collection of toy models for asymptotic safety that apply to QFTs in low dimensions (see, e.g., Refs.\ \cite{Gies:2009hq,Braun:2010tt}). In our example, we focus on the question how both the UV and the IR limit of the RG trajectory are determined by interacting fixed points with different degrees of symmetry.

\section{Effective action functional for multi-field models}
\label{Sec:Effective action functional for multi-field models}

Building on previous work \cite{PhysRevE.90.052129}, we derive the functional RG $\beta$-functions from the nonperturbative flow equation for the scale-dependent effective action functional $\Gamma = \Gamma\left[\{ \phi_i \}\right]$ in $d$-dimensional Euclidean space \cite{Wetterich:1992yh} (see, e.g., \cite{Berges:2000ew,Polonyi:2001se,Pawlowski:2005xe,Gies:2006wv,Delamotte:2007pf} for reviews).
Our starting point is an \textit{ansatz} for $\Gamma$ to leading order in the derivative expansion
\be
\Gamma =\int \textrm{d}^dx \left[ \frac{1}{2}\sum_{i = 1}^n Z_{i}\, (\partial \phi_i)^2 + U\!\left(\{ \phi_j \}\right) \right] ,
\label{Eq:EffectiveAction}
\ee
where the summation runs over $n$ distinct field degrees of freedom $\phi_i$ (defined in the $N_i$-dimensional vector space representation of the $O(N_i)$ symmetry group). The renormalization factors $Z_i$ and the effective potential $U$ are both assumed to be scale-dependent.\footnote{Here, we include terms to $\mathcal{O}(\partial^2)$ and neglect a possible field dependence of the scale-dependent renormalization factors $Z_{i}$ (that are evaluated at the minimum of the effective potential $U$). Terms of the type $\sim \big(\partial \phi_i^2\big)^2$ that in principle contribute at the same order in the derivative expansion are not taken into account.}

  In the following, we exploit the symmetry of the model and write $U$ in terms of the field invariants $\rho_i = \phi_i^2 / 2$. Furthermore, we choose to expand the effective potential to some finite order $M\geq 2$ around a possibly nonvanishing scale-dependent minimum $\kappa_i$, where \mbox{$\left. \partial U_k / \partial \rho_i \,\right|_{\{\rho_j = \kappa_j\}} = 0$} and \mbox{$\left. \partial^2 U_k / \partial \rho_i^2 \,\right|_{\{\rho_j = \kappa_j\}} > 0$}. We have
\be
\hspace{-5pt} U = \sum_{m_1 + \ldots + m_n = 2}^M \lambda_{m_1 \cdots\, m_n} \frac{\prod_{i = 1}^n \left( \rho_i - \kappa_i \right)^{m_i}}{\prod_{i=1}^{n} m_i!} ,
\label{Eq:EffectivePotential}
\ee
where $\lambda_{m_1 \cdots\, m_n}$ are the scale-dependent couplings. The above expansion \eqref{Eq:EffectivePotential} effectively introduces a large number of couplings (at each value of the RG scale parameter, \mbox{$0 \leq k < \Lambda$}), but only a few of them appear in the \textit{bare} action $S = \Gamma(k=\Lambda)$ defined at scale $\Lambda$.

The $\beta$-functions to one-loop order in the $\epsilon$-expansion may also be obtained from the nonperturbative RG flow equation. This is achieved by employing an expansion around the upper critical dimension and restricting the functional space to those operators that appear in the bare action.\footnote{By virtue of one-loop universality, the $\mathcal{O}(\epsilon)$-expanded functional RG $\beta$-functions are exact and independent of the chosen (nonperturbative) regulator.}
In the case of two fields these $\beta$-functions agree with those given in Ref.\ \cite{Folk:2008mi}, as expected by one-loop universality. In the following, we consider a model in $d = 3$ dimensions with three different field degrees of freedom, $\phi_1, \phi_2$, and $\phi_3$, with $N_1$, $N_2$, and $N_3$ field components, respectively, and compare our results \cite{PhysRevE.90.052129} explicitly with the Wilsonian momentum-shell RG to one-loop order in the $\epsilon$-expansion. As outlined in Sec.\ \ref{Sec:Introduction}, our main goal is to complement the functional RG with the $\epsilon$-expansion to identify and characterize all possible multicritical scaling solutions relevant in the IR scaling regime. A similar strategy was also chosen in Ref.\ \cite{O'Dwyer:2007ia}, where the Polchinski version of the nonperturbative RG \cite{Polchinski:1983gv} was contrasted to the $\epsilon$-expansion to investigate multicritical points for a scalar theory with a single order parameter.

\section{RG fixed points to $\mathcal{O}(\epsilon)$ in the three-field model}
\label{Sec:RG fixed points in the three-field model}

\begin{figure*}[!t]
\includegraphics[width=0.29\linewidth]{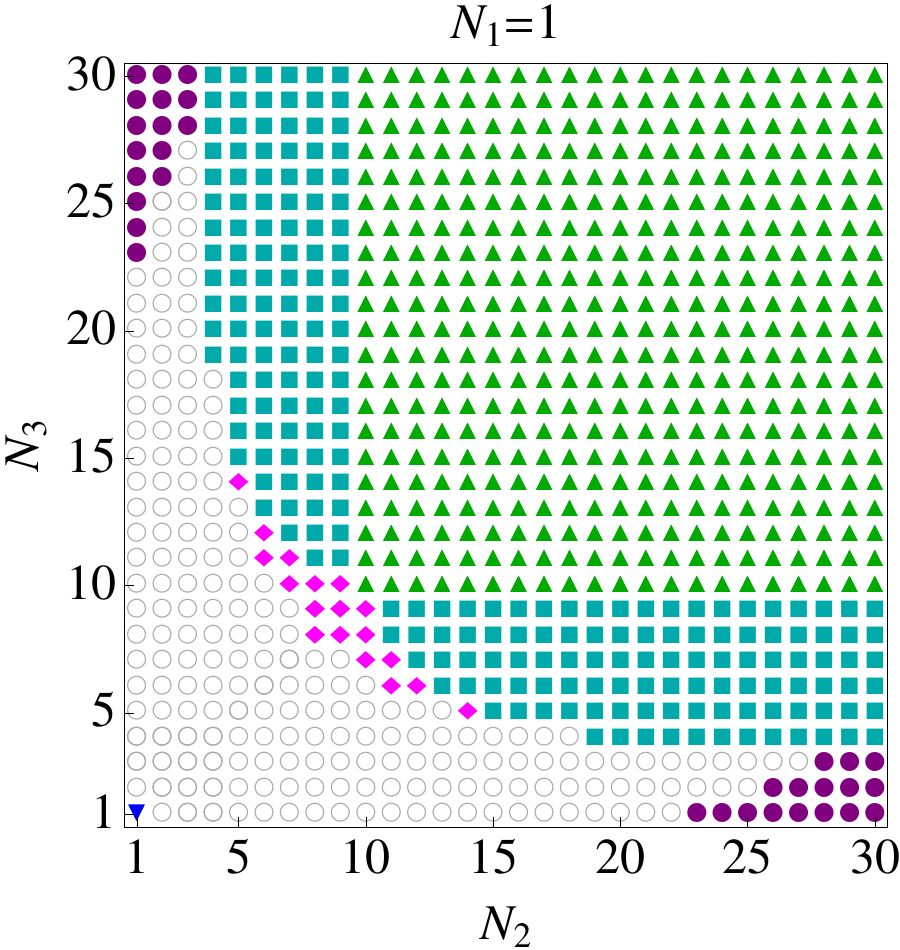}
\hspace{60pt}
\includegraphics[width=0.29\linewidth]{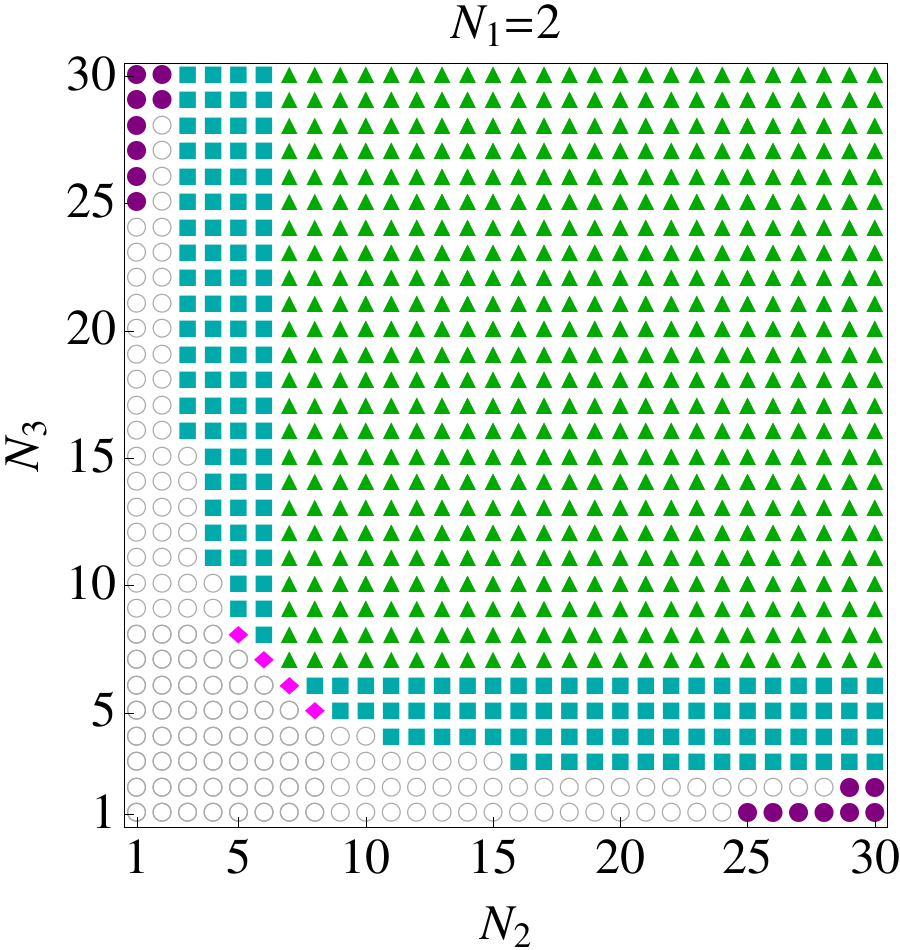}
\caption{\label{fig1}IR-stable and marginally-stable FPs for the three-field model in the one-loop $\epsilon$-expansion ($\epsilon = 1$) in the $(N_2, N_3)$-plane, where $N_1 = 1$ (\textbf{left}) and $N_1 = 2$ (\textbf{right}). We identify the IFP ({\small{\color{blue}$\blacktriangledown$}}), the DFP ({\small {\color{darkgreen}$\blacktriangle$}}), the DIFP ({\color{purple}$\bullet$}), and the DBFP ({\tiny{\color{cyan}$\blacksquare$}}). In addition to these scaling solutions, we find another IR-stable fully-coupled FP, the ACFP (\protect\rotatebox[origin=c]{45}{\color{pink}\tiny$\blacksquare$}), while no stable FP is found for small values of $N_2$ and $N_3$ ({\color{gray}$\circ$}).}
\end{figure*}

Employing the Wilsonian momentum-shell RG to one-loop order in the $\epsilon$-expansion, we obtain the following set of $\beta$-functions:\footnote{We introduce the following notation for the $\beta$-functions
\be
\beta_{m_1 \cdots\, m_n} \equiv k \frac{\partial \lambda_{m_1 \cdots\, m_n}}{\partial k} ,
\ee
which are expressed in terms of the rescaled couplings
\be
\lambda_{m_1 \cdots\, m_n} \rightarrow K_d k^{-d}\left( \prod_{i = 1}^n Z_i^{-m_i} k^{(d-2) m_i} \right) \lambda_{m_1 \cdots\, m_n} ,
\ee
where $K_d = \left[ (4\pi)^{d/2} \Gamma(d/2 + 1) / 2 \right]^{-1}$ and the RG scale is given by $k = e^{s} \Lambda$, $-\infty < s \leq 0$.}
\bea
&& \hspace{-10pt} \beta_{200} = - \epsilon \lambda_{200} + (N_1+8) \lambda_{200}^2 + N_2 \lambda_{110}^2 + N_3 \lambda_{101}^2 , \label{Eq:BetaEpsilon1} \\
&& \hspace{-10pt} \beta_{020} = -\epsilon \lambda_{020} + N_1 \lambda_{110}^2 + (N_2+8) \lambda_{020}^2 + N_3 \lambda_{011}^2 , \\
&& \hspace{-10pt} \beta_{002} = -\epsilon \lambda_{002} + N_1 \lambda_{101}^2 + N_2\lambda_{011}^2 + (N_3+8) \lambda_{002}^2 , \\
&& \hspace{-10pt} \beta_{110} = - \epsilon \lambda_{110} + (N_1 + 2) \lambda_{110} \lambda_{200} + (N_2 + 2) \lambda_{020} \lambda_{110} \nonumber\\
&& \hspace{22pt} + \: N_3 \lambda_{011} \lambda_{101} + 4 \lambda_{110}^2 ,\\
&& \hspace{-10pt} \beta_{101} = - \epsilon \lambda_{101} + (N_1 + 2) \lambda_{101} \lambda_{200} + N_2 \lambda_{110}\lambda_{011} \nonumber\\
&& \hspace{22pt} + \: (N_3 + 2) \lambda_{002} \lambda_{101} + 4 \lambda_{101}^2 ,\\
&& \hspace{-10pt} \beta_{011} = - \epsilon \lambda_{011} + N_1 \lambda_{101}\lambda_{110} + (N_2 + 2) \lambda_{011} \lambda_{020} \nonumber\\
&& \hspace{22pt} + \: (N_3 + 2) \lambda_{002} \lambda_{011} + 4 \lambda_{011}^2 .\label{Eq:BetaEpsilon6}
\eea

Let us briefly highlight the difference of the above $\beta$-functions to those derived within the functional RG approach \cite{PhysRevE.90.052129}: 
In the latter case higher-order interactions (generated by the RG flow towards the IR) are explicitly taken into account.
Therefore, the $\beta$-functions for the quartic couplings receive contributions from higher-order couplings (their scale-dependence being characterized by additional $\beta$-functions that are determined within this approach).
Moreover, the functional RG represents a \textit{massive} renormalization scheme, and accordingly, mass parameters explicitly enter all $\beta$-functions, resulting in their non-polynomial structure.
Details can be found in \mbox{Ref.\ \cite{PhysRevE.90.052129, Eichhorn:2013zza}}.

We solve for the zeros of the $\beta$-functions, Eqs.\ \eqref{Eq:BetaEpsilon1} -- \eqref{Eq:BetaEpsilon6} to obtain the FPs of the RG flow. 
Their stability is captured by the (critical) scaling spectrum, defined in terms of the eigenvalues of the stability matrix at the FP:
\be
\theta \in - \spec \left( \frac{\partial \beta_{M}}{\partial \lambda_{M'}} \right)_{\textrm{FP}},
\label{Eq:StabilityMatrix}
\ee
where $\beta_M \equiv \beta_{m_1 \cdots\, m_n}$ and $\lambda_{M'} \equiv \lambda_{m'_1 \cdots\, m'_n}$. We refer to a FP as IR-stable, if all eigenvalues \eqref{Eq:StabilityMatrix} are negative. Mass-like perturbations in the bare action are \textit{always relevant}.
It is the parameters corresponding to these mass-like operators that need to be tuned in order to reach the IR scaling solution (assuming that the microscopic parameters of the model are in the domain of attraction of that particular FP).\footnote{To comply with the notation introduced in \cite{PhysRevE.90.052129}, we will assume that the eigenvalues are labeled in descending order, \textit{i.e.}, $\theta_1 \geq \theta_2 \geq \ldots \,$, while $\theta_{\mu} > 0$, for $1\leq \mu \leq n$, corresponding to mass-like perturbations (where $n = 3$ for the three-field model).} If no stable FP is found, additional fine-tuning might be necessary to observe a continuous phase transition with universal scaling exponents. However, usually this is experimentally unfeasible, and we therefore conjecture that the corresponding systems will not feature multicritical behavior. In such a case the RG flow trajectories diverge and we expect that if a phase transition is observed it will be of first order.
Nevertheless, the divergence of RG trajectories might manifest itself only deep in the IR. E.g., the RG trajectory of a theory for which the couplings at the UV scale $\Lambda$ are chosen to be close to a particular symmetry-enhanced subspace (where another fixed point with four relevant directions exists) might display a weak scale-dependence -- a very slow walking -- over a wide range of scales. For all practical purposes, such a scenario is difficult to distinguish from conformal scaling behavior.

Within the one-loop $\epsilon$-expansion our study uncovers the following expected scaling solutions:
\begin{itemize}
  \item \textit{Isotropic fixed point (IFP)}, featuring a symmetry enhancement to an $O(N_1+N_2+N_3)$ symmetry and consequently has coordinates $\lambda_{200} = \lambda_{020} = \lambda_{002} = \lambda_{110} = \lambda_{101} = \lambda_{011}$; 
  \item \textit{Decoupled fixed point (DFP)} is characterized by a complete decoupling of all sectors, \textit{i.e.}, $\lambda_{110} = \lambda_{101} = \lambda_{011} = 0$;
  \item \textit{Decoupled isotropic fixed points (DIFP)} are characterized by a partial decoupling of the sectors as well as partial symmetry enhancement. One representative in this class is given by $\lambda_{002} = \lambda_{020} = \lambda_{011}$, while $\lambda_{101} = \lambda_{110} = 0$, and features an enhanced $O(N_2+N_3)$ symmetry;
  \item \textit{Decoupled biconical fixed points (DBFP)} are partially decoupled but feature no symmetry enhancement, e.g., we might have $\lambda_{101} = \lambda_{110} = 0$ and $\lambda_{011} \neq \lambda_{020}$.
\end{itemize}
These FPs were previously identified in the framework of the functional RG in Ref.\ \cite{PhysRevE.90.052129} where their scaling and stability properties were discussed in detail.
Within the $\epsilon$-expansion, we confirm our previous finding that the three-field models in \mbox{$d = 3$} dimensions exhibit regions in the space of field components $N_i$ where no IR-stable FP exists; \mbox{cf.\ Fig.\ \ref{fig1}}. Specifically, this implies that particular three-field models with a given set of $(N_1, N_2, N_3)$ do not feature multicritical behavior without additional fine-tuning.
A similar absence of IR-stable multicritical FPs was observed in Ref.\ \cite{cla15} where the effect of competing order was investigated on fermionic quantum criticality (see also Ref.\ \cite{She:2010cj}).
%


\subsection*{Fully-coupled FPs}

To uncover additional IR-stable FPs, we inspect the scaling solutions as a function of the parameters $N_i$, \mbox{see Fig.\ \ref{fig1}.}

\subsubsection{Asymmetrically-coupled FP}

Our main result is the discovery of a new FP, which is completely coupled, \textit{i.e.}, $\lambda_{101} \neq 0$, $\lambda_{110}\neq 0$, and $\lambda_{011} \neq 0$, but does not feature any symmetry enhancement.
In the following, we will refer to this scaling solution as the \textit{asymmetrically coupled fixed point (ACFP)}.
It defines a genuine new universality class that cannot be obtained as a generalization of the Wilson-Fisher FP, and occurs for the first time in the three-field model. This new universality class relies crucially on the presence of three competing orders, and cannot occur in systems with a smaller number of order parameters.

\begin{table}[!b]
  \setlength{\tabcolsep}{4pt}
  \renewcommand{\arraystretch}{1.3}
  \begin{tabular}{ccc|cccccc}
$N_1$ & $N_2$ & $N_3$ & $\lambda_{200}$& $\lambda_{020}$& $\lambda_{002}$& $\lambda_{110}$& $\lambda_{101}$& $\lambda_{011}$ \\ \hline \hline
1 & 8 & 8 & 0.105 &0.062 & 0.062 & 0.020& 0.020 & -0.002 \\\hline\hline
1& 8 & 9 & 0.109 & 0.062 & 0.059& 0.013 & 0.008 & -$4\cdot 10^{-4}$ \\ \hline \hline
1& 8 & 10 & 0.110 & 0.063 & 0.056 & 0.012 & 0.002 & -$7 \cdot 10^{-5}$ \\ \hline \hline 
\end{tabular}\\
  \vspace{0.1cm}
  \setlength{\tabcolsep}{3.7pt}
\begin{tabular}{ccc|cccccc}
$N_1$ & $N_2$ & $N_3$ & $\theta_4$ & $\theta_5$ & $\theta_6$ & $\theta_7$ & $\theta_8$ & $\theta_9$\\ \hline \hline
1 & 8 & 8 & 0 & -0.096 & -0.255 & -0.996 & -0.996 & -1.000 \\\hline\hline
1 & 8 & 9 & -0.019 & -0.046 & -0.274 & -0.977 & -0.999 & -1.000 \\ \hline\hline
1& 8 & 10 & -0.007 & -0.037 & -0.294 & -0.985 & -1.000 & -1.000 \\ \hline \hline
\end{tabular}
\caption{\label{tab1}As we decrease the number of field components $N_3$ and pass through the point $N_1 = 1$, $N_2 = N_3 = 8$ ($\epsilon = 1$), the fully-coupled FP ceases to be IR-stable -- the scaling exponent $\theta_4$ becomes non-negative as the value of $N_3$ is lowered.}
\end{table}
To illustrate its properties, we give the corresponding values of the dimensionless, renormalized couplings $\lambda_{m_1 m_2 m_3}$ and the critical exponents $\theta_4 , \theta_5,  \ldots$ at selected points in the $(N_2, N_3)$-plane, for $N_1 = 1$; see \mbox{Tab.\ \ref{tab1}}. Mass parameters do not appear in the $\beta$-functions to the given order of the $\epsilon$-expansion. Therefore, the three relevant scaling exponents $\theta_1$, $\theta_2$, and $\theta_3$ are not provided in the following.

\begin{figure}[!t]
\includegraphics[width=0.8\linewidth]{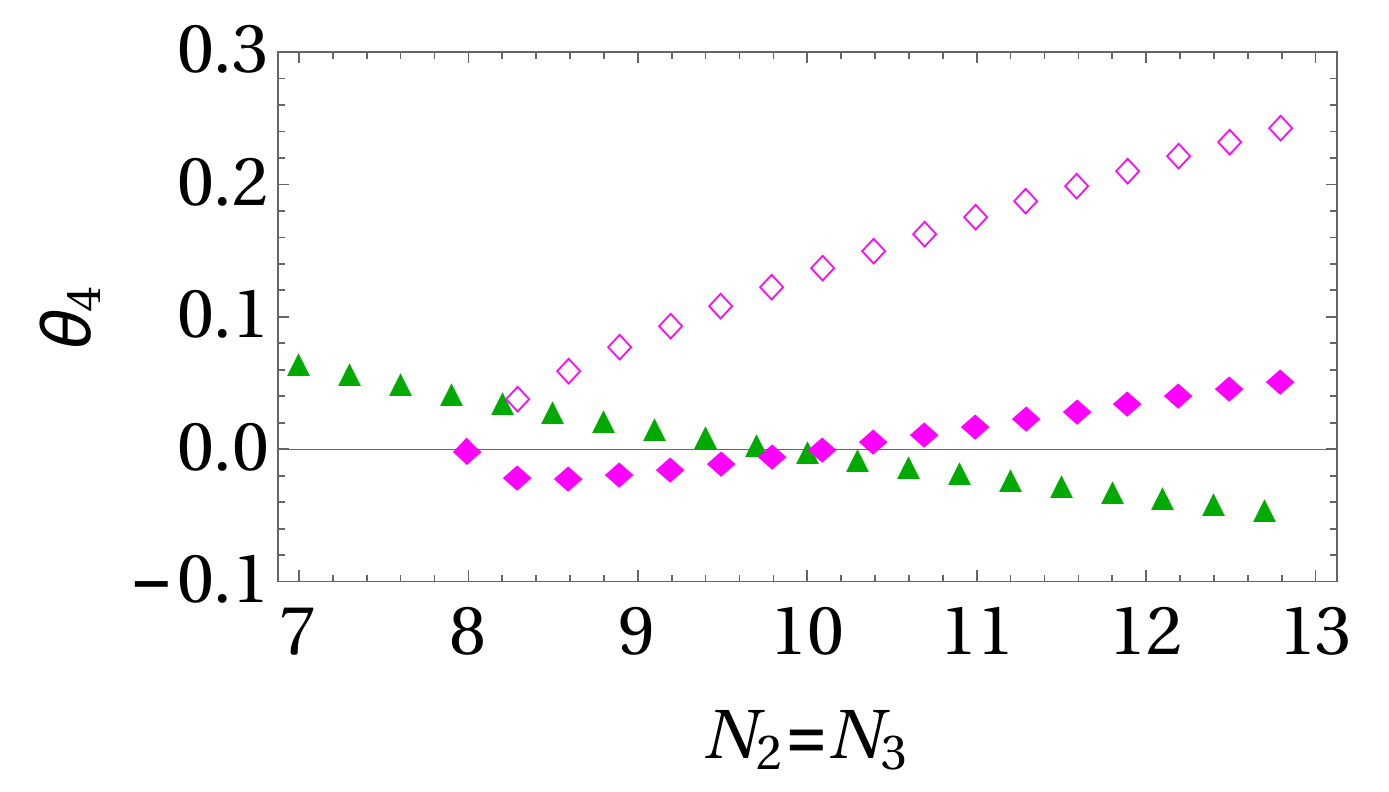}
\caption{\label{fig2}Varying the number of field components $N \equiv N_2 = N_3$, while keeping $N_1 = 1$ fixed, we observe that the asymmetric fully-coupled FP (\protect\rotatebox[origin=c]{45}{\color{pink}\tiny$\blacksquare$}) appears together with a second fully-coupled (unstable) FP (\protect\rotatebox[origin=c]{45}{\color{pink}\tiny$\square$}). By inspecting the sign of the exponent $\theta_4$, that decides about stability of the considered scaling solution, we conclude that the ACFP is IR-stable until $N = 10$ where it collides with the DFP ({\small {\color{darkgreen}$\blacktriangle$}}). From there the DFP takes over stability.}
\end{figure}

We follow the fully-coupled asymmetric FP along the $N\equiv N_2 = N_3$ direction, where we expect that it should collide with the DFP at some critical value of $N$; cf.\ \mbox{Fig.\ \ref{fig2}}. In fact, we find that the two FPs exchange their stability properties at ($N = 10$, $N_1 = 1$). That is, at the collision point the exponent that decides about the stability properties of the scaling solution, $\theta_4$, changes its sign for each of the two solutions. If we attempt to continue the asymmetric FP to smaller values of $N$, we observe that it disappears into the complex plane at $N = 8$, $N_1 = 1$, together with another fully-coupled FP which is always unstable -- both FPs become \emph{inaccessible} for small values of $N_i$. From these results one might conclude that the fully-coupled asymmetric FP will not be of any significance experimentally: The one-loop $\epsilon$-expansion seems to suggest that there is a threshold value $N_{i} \simeq 5$, for all $i = 1,2,3$, below which the ACFP disappears completely (cf.\ Fig.\ \ref{fig1}). We show in  Sec.\ \ref{Sec:FunctionalRG}  that the functional RG provides a quantitatively more reliable estimate for the critical values of $N_{i}$.

It is interesting to note that the spontaneous creation/annihilation of two FPs -- at least one of which is IR-stable and therefore might be relevant for the multicritical scaling behavior in the IR -- does not appear in the $O(N_1)\oplus O(N_2)$ models. In the case of two coupled order parameters, we may associate exactly \emph{one} IR-stable FP to each pair of values $(N_1, N_2)$. Such a scaling solution can then be continued to \emph{all} values of $N_i$, but in that process might lose its stability to another FP. That is, \emph{in principle} each of the possible multicritical universality classes is accessible for all values of field components $N_1$ and $N_2$ via additional fine-tuning of parameters. This is in sharp contrast to the three- and four-field models (see Sec.\ \ref{Sec:Four-field model to one-loop order}).

\subsubsection{Generalized BFP and regions without IR stable FP}

In general, the space of renormalized couplings $\lambda_{m_1 \,\cdots\, m_n}$ features closed subspaces that are characterized by enhanced symmetries and the decoupling phenomenon: Whenever one of the sectors decouples, and the couplings between sectors vanish, fluctuations cannot regenerate the mixed couplings, and therefore the RG flow stays within that space, making it an RG-invariant subspace.
With the discovery of the new asymmetrically coupled FP we may complete this picture in the following way: We may state that each of these subspaces (excluding its symmetry-enhanced or decoupled subspaces) contains at least one FP. In fact, from our analysis we find that almost all of these subspaces will feature an \emph{IR-stable} FP for a particular set of values $N_i$, with one notable exception -- the BIFP, cf.\ Appendix \ref{Sec:Appendix}. This scaling solution is associated to the partially symmetry-enhanced subspace and is nowhere stable.

While in principle such a universality class exists, it would require a higher degree of fine-tuning to reach it. 
Thus, the associated pattern of symmetry-breaking is not expected to be relevant experimentally. 
The BIFP would be a natural candidate to take over stability from the IFP as soon as it becomes unstable, just as the BFP takes over stability from the IFP in the two-field-model. 
The additional relevant directions of the BIFP prevent this scenario from being realized, and imply that the three-field case features a region in the space of the $N_i$ that is devoid of stable FPs. 

\subsection*{Multi-field theories as toy models for asymptotic safety and IR-completeness}
  In order for a QFT to provide a viable description of a set of degrees of freedom and their interactions on all scales, \textit{i.e.}, in order for the theory to be fundamental, it must reach a renormalization group FP in the UV and IR, respectively.\footnote{In principle, more exotic scenarios as, e.g., limit cycles, might also be viable.} 
  Here, we provide a set of models that features a large number of complete trajectories -- that run into nontrivial FPs both in the UV and IR.

In this context, it is important to realize that in principle a given FP can be reached asymptotically in either one of the two limits,  if it features at least one critical exponent that differs in sign from the others. If a FP should be reached in the UV, all irrelevant couplings need to be tuned in such a way that the RG trajectory lies within the UV-critical hypersurface of the FP. In the context of high-energy physics, this implies that the values of all irrelevant couplings correspond to \emph{predictions} of the model, \textit{i.e.}, for the model to be asymptotically safe, there is exactly one possible value for each irrelevant coupling. On the other hand, if the FP is reached in the IR, the renormalization group flow is automatically drawn towards it along the irrelevant directions, and it is the relevant directions that require tuning.

The large number of interacting FPs in our model provides a variety of different complete trajectories, connecting pairs of nontrivial FPs --subject to global properties of the flow. Furthermore, due to the possibility of symmetry enhancement at FPs, multi-field models are also of interest from the point of view of fundamental physics. For instance, it has been conjectured that quantum gravity should exhibit a violation of Lorentz symmetry, connected to anisotropic scaling in the ultraviolet \cite{Horava:2009uw,DOdorico:2014iha}. As violations of Lorentz symmetry are strongly constrained in the IR, such a setting requires a rather precise restoration of Lorentz symmetry at small scales. Here, we identify a set of models (defined by the number of field components in the different field sectors, $N_i$) where a symmetry enhancement -- in our case an enhanced rotational symmetry in field space -- requires additional tuning. That is, the symmetry enhancement scenario is thus considered ``unnatural''. Interestingly, we also observe a number of realizations of these theories, in which an enhancement of symmetry is the most natural IR-endpoint of a trajectory, as all other existing FPs require a higher degree of tuning in order to reach them.  Note that some degree of tuning is always required, as no fixed point comes with only IR-attractive directions. Demanding that the trajectory ends in an IR fixed point thus requires tuning at least three parameters. To avoid symmetry-enhancement in these cases requires \emph{additional} tuning.

\begin{figure}[!t]
\includegraphics[width=0.75\columnwidth]{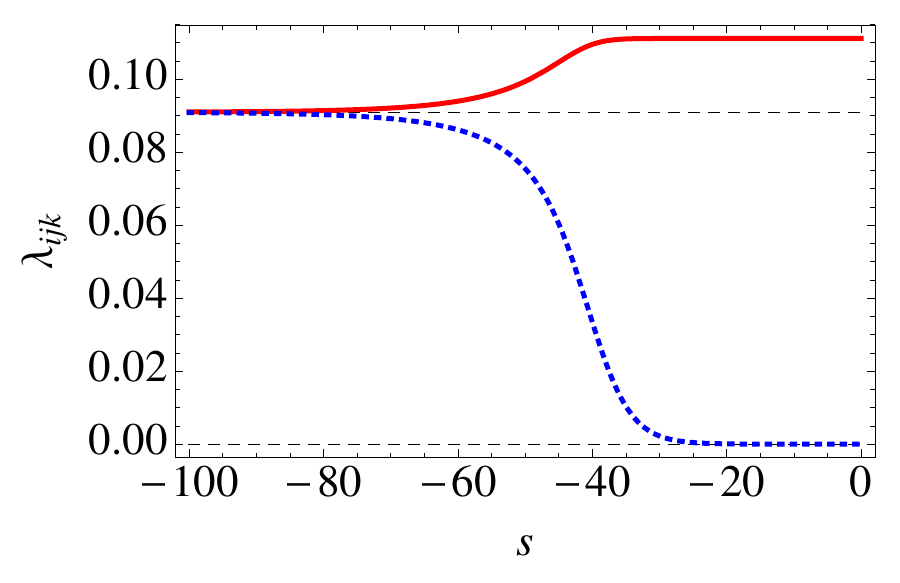}\newline
\includegraphics[width=0.75\columnwidth]{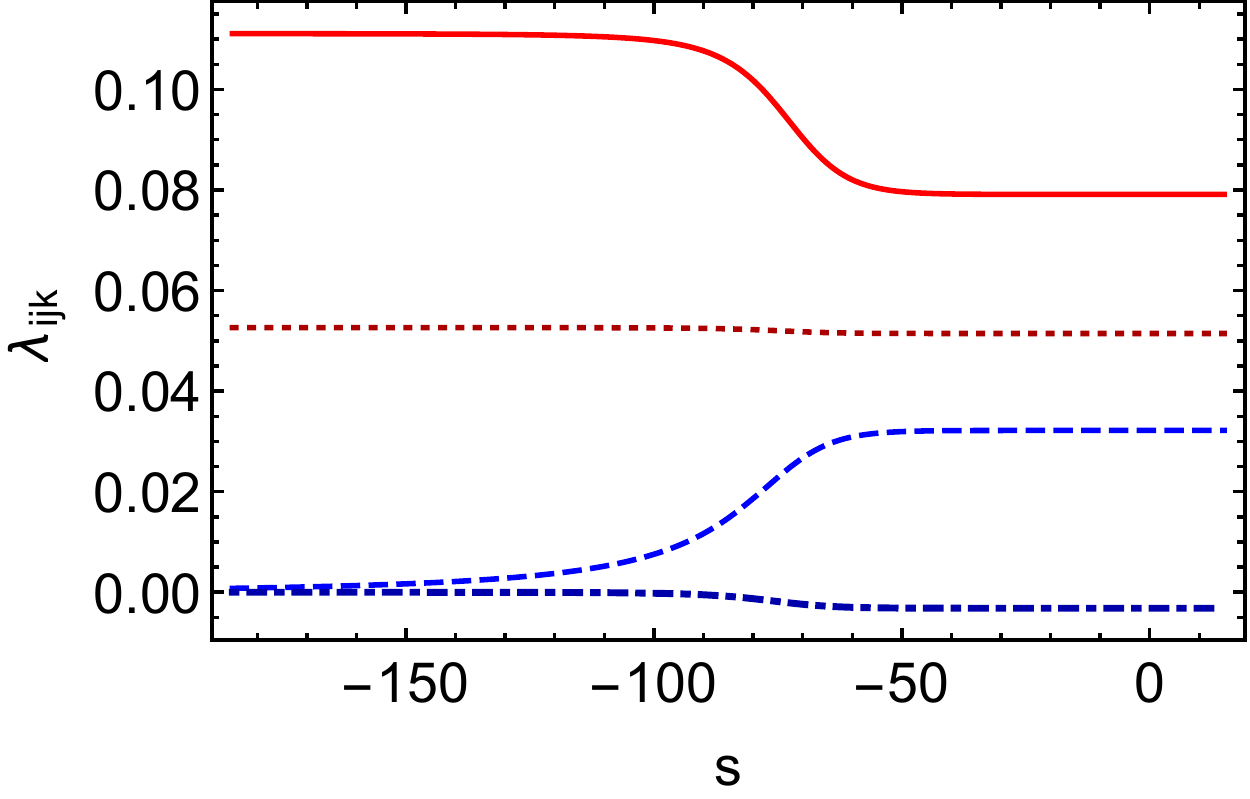}\newline
\caption{\label{fig6}Upper panel: We plot RG flow trajectories connecting the DFP in the UV (large positive $s$) with the symmetry-enhanced IFP in the IR (large negative $s$). We show the couplings $\lambda_{200}=\lambda_{020}=\lambda_{002}$ (solid red line) and $\lambda_{110}=\lambda_{101}=\lambda_{011}$ (dotted blue line) for $N_1=N_2=N_3 = \boldsymbol{1}$ as a function of the RG scale $s=\ln(k/\Lambda)$. In this case the IFP is the stable FP and the enhanced symmetry emerges naturally as a result of the flow towards the IR. Lower panel: We show a trajectory that connects the ACFP in the UV to the DFP in the IR for the case $N_1=1, N_2=N_3=11$, for which the DFP is IR stable, and the ACFP has one IR-relevant direction, which triggers the flow to the IR. We show the couplings $\lambda_{200}$ (solid red line), $\lambda_{020}=\lambda_{002}$ (dotted dark red line), $\lambda_{110}=\lambda_{101}$ (dashed blue line) and $\lambda_{011}$ (dot-dashed dark blue line).
}
\end{figure}

\begin{table*}[!t]
  \setlength{\tabcolsep}{7.7pt}
  \renewcommand{\arraystretch}{1.3}
  \begin{tabular}{cccc|cccccccccc}
$N_1$ & $N_2$ & $N_3$& $N_4$ & $\lambda_{2000}$& $\lambda_{0200}$& $\lambda_{0020}$& $\lambda_{0002}$&$\lambda_{1100}$& $\lambda_{1010}$& $\lambda_{0110}$& $\lambda_{1001}$&$\lambda_{0101}$&$\lambda_{0011}$\\ \hline \hline
1 & 9 & 9 & 9 & 0.110 & 0.059 & 0.059 & 0.059 & 0.007 & 0.007 & -$1\cdot10^{-4}$ & 0.007 & -$1\cdot10^{-4}$ & -$1\cdot10^{-4}$ \\ \hline \hline
1 & 10 & 9 & 9 & 0.110 & 0.056 & 0.059 & 0.059 & 0.001 & 0.006 & -$2\cdot10^{-5}$ & 0.006 & -$2\cdot10^{-5}$ & -$1\cdot10^{-4}$ \\ \hline\hline
1 & 10 & 9 & 8 & 0.109 & 0.055 & 0.059 & 0.062 & 0.003 & 0.008 & -$8\cdot10^{-5}$ & 0.013 & -$1\cdot10^{-5}$ & -$3\cdot10^{-4}$\\ \hline \hline 
  \end{tabular}\\
  \vspace{0.1cm}\setlength{\tabcolsep}{7.9pt}
  \begin{tabular}{cccc|cccccccccc}
$N_1$ & $N_2$ & $N_3$& $N_4$ & $\theta_5$ & $\theta_6$ & $\theta_7$ & $\theta_8$ & $\theta_9$ & $\theta_{10}$& $\theta_{11}$ & $\theta_{12}$& $\theta_{13}$& $\theta_{14}$ \\\hline \hline
1 & 9 & 9 & 9 & -0.012 & -0.028 & -0.028 & -0.295 & -0.295 & -0.295 & -0.986 & -1.000 & -1.000 & -1.000 \\ \hline\hline
1 & 10 & 9 & 9 & -0.004 & -0.016 & -0.024 & -0.295 & -0.314 & -0.314 &-0.993 & -1.000 & -1.000 & -1.000 \\ \hline\hline 
1 & 10 & 9 & 8 &  -0.009 & -0.022 & -0.046 & -0.294 & -0.306 & -0.314 & -0.975 & -0.999 & -1.000 & -1.000  \\ \hline \hline
  \end{tabular}
  \caption{\label{tab2}{Stable, fully coupled FP in the four-field case. There are four relevant directions; the corresponding exponents $\theta_1$, \ldots, $\theta_4$ are not provided.}}
\end{table*}

In particular, we will focus on two examples:
  The first involving the ACFP as a UV fixed point, thus defining a toy-model for an asymptotically safe model. Here we pick $N_1=1,N_2=N_3=11$, where the ACFP has one IR-relevant direction; this triggers a flow to the DFP in the IR.
As a second example, we consider a region of $N_i$ where, e.g., the IFP is stable (with three relevant directions) it is a natural candidate FP for RG trajectories in the IR, see Fig.~\ref{fig6}. Thus, a model that has been rendered asymptotically safe, e.g., by defining it at the DFP, can only be infrared complete when at least three directions are tuned and it is the symmetry-enhanced   IFP which provides the lowest number of relevant directions. In this case, IR FPs with a lower degree of symmetry will typically require a higher degree of fine tuning.


\section{Four-field model to one-loop order in the $\epsilon$-expansion}
\label{Sec:Four-field model to one-loop order}

We proceed in an analogous manner for the four-field model.
Our main goal here is to confirm that the two novel features of the class of $O(N_1)\oplus O(N_2)\oplus O(N_3)$-field models -- the possible existence of theories without an IR-stable FP and the existence of a new fully coupled FP -- carry over to the case of larger numbers of fields. 
The $\beta$-functions are given by the obvious generalization of Eqs.\ \eqref{Eq:BetaEpsilon1} -- \eqref{Eq:BetaEpsilon6} to the case where one additional field degree of freedom with $O(N_4)$ symmetry is added.
Determining the zeros of the beta functions, we find that a new FP which is fully coupled and does not feature any symmetry enhancement, exists and is stable at selected points in the space of the $N_i$, \textit{i.e.}, a FP that appears for the first time in the four-field case similar to the role of the ACFP in the three-field case, cf.\ Tab.\ \ref{tab2}. 
In this context, stability is of course defined as the existence of no more than four relevant directions.

The new FP collides with the DFP at $N_i=10$ and becomes unstable.
Moreover, our results indicate that no FP is stable, e.g., at the point $N_1=1$, $N_2=N_3=N_4=8$. 
Together with the results in Tab.\ \ref{tab2}, this suggests that a structurally similar picture to the three-field case carries over to the four-field case: 
The IFP will be stable for very small values of the $N_i$, before it is destabilized. 
Keeping $N_1=1$ fixed and increasing $N_2=N_3=N_4$, we pass through a regime without a stable FP, \textit{i.e.}, with nonuniversal behavior only. 
At $N_i=9$, the new FP then appears from the complex plane, and is stable until it collides with the DFP, that takes over stability for all larger values of the $N_i$. 
Based on our findings in the three-field model, we expect that our $\mathcal{O}(\epsilon)$ estimates for the $N_i$, at which FPs are stable, are considerably larger than the correct values. 
As can be tested within, e.g., the LPA 4, the functional RG is more reliable when it comes to quantitative estimates.

Based on our findings in the three- and four-field case, we therefore conjecture that models with larger numbers of competing orders will not feature multicritical behavior without additional fine-tuning.
%


\section{Results from the functional RG}
\label{Sec:FunctionalRG}

Details on the derivation of the functional RG equations for the model can be found in Ref.\ \cite{PhysRevE.90.052129}.
As an important difference to the results of the $\epsilon$-expansion we note that the critical values of $N_i$, at which the DFP becomes stable, are found to lie at much lower values even in the simplest possible truncation of the functional RG equations -- the local potential approximation (LPA). It is defined by an expansion of the scale-dependent effective potential \eqref{Eq:EffectivePotential} in terms of point-like interactions, \emph{without} taking into account the scale-dependence of the renormalization factors, \textit{i.e.}, $\eta_i \equiv - k \partial \ln Z_i / \partial k = 0$ (see Eq.\ \eqref{Eq:EffectiveAction} for the definition of the parameters and couplings in the given model).
The LPA-type truncation of the functional RG provides a quantitatively more precise estimate than the one obtained from the one-loop $\epsilon$-expansion, which may be confirmed independently by employing nonperturbative scaling relations (see \mbox{Ref.\ \cite{2015PhRvE..91f2112B}}).
\begin{figure}[!t]
\includegraphics[width=0.8\linewidth]{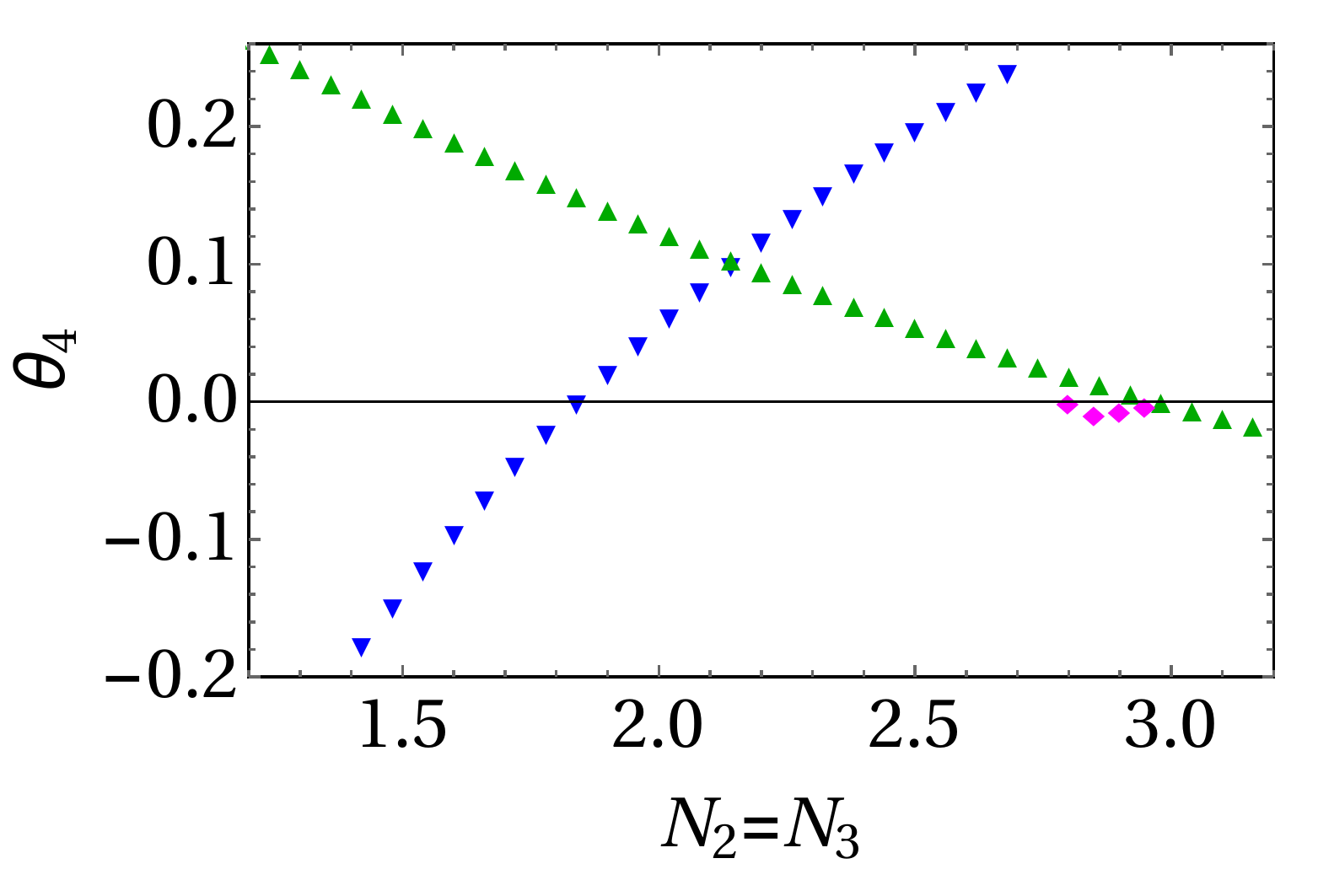}
\caption{\label{fig3}We show the values for the critical exponent $\theta_4$ obtained from the functional RG (LPA 4 + $\eta$) as a function of $N \equiv N_2 = N_3$ ($N_1 = 1$, $d = 3$), for the IFP ({\small{\color{blue}$\blacktriangledown$}}), the DFP ({\small {\color{darkgreen}$\blacktriangle$}}), and the new coupled asymmetric FP (\protect\rotatebox[origin=c]{45}{\color{pink}\tiny$\blacksquare$}). We find the same picture as in the $\epsilon$-expansion: The ACFP is IR-stable in a small region $2.8 \lesssim N \lesssim 3$ before the DFP takes over stability. The upper and lower boundary values to this region are significantly smaller than those obtained from the one-loop $\epsilon$-expansion ($\epsilon = 1$): $8 \leq N \leq 10$ (cf.\ Fig.\ \ref{fig2}).}
\end{figure}

Employing an LPA truncation to 4th order in the fields and including the scale-dependence of the renormalization factors, \textit{i.e.}, $\eta_i \neq 0$ (which we will refer to as LPA $4 + \eta$ in the following), we confirm the qualitative behavior of the $\epsilon$-expansion: 
Fixing $N_1 = 1$ and increasing $N \equiv N_2=N_3$, the IFP becomes unstable around $N \simeq 1.85$. 
For larger values of $N$ there is no IR-stable FP, until the new ACFP appears and becomes stable; cf.\ Fig.\ \ref{fig3}.
In contrast to the $\epsilon$-expansion, this already happens at $N \simeq 2.8$. 
Finally, at $N \simeq 3$ the asymmetrically coupled FP exchanges its stability with the DFP, which remains IR-stable for all $N \gtrsim 3$. 
The mechanism by which the new ACFP appears is completely analogous to the situation observed in the one-loop $\epsilon$-expansion: It appears from the complex plane together with another FP and immediately takes over stability. Note however, that the region of values $N$ where the ACFP is stable is shifted to significantly smaller values of $N$ bringing it into the reach of physically interesting models. Thus, the asymmetrically coupled FP might actually be of interest for efforts to establish the phase diagram of strongly-correlated many-body systems either experimentally or via lattice Monte Carlo techniques, for an overview, see, e.g., Ref.~\cite{Vicari:2007ma}. Using scaling relations to estimate the stability regime for the DFP, we find that the LPA $4 + \eta$ slightly overestimates the width of the region where the ACFP is stable.
We expect that the region where the ACFP is stable, becomes even smaller at higher orders of the LPA. 
In fact, this might account for the fact that it was not discovered in our previous analysis \cite{PhysRevE.90.052129} based on a LPA to 8th order in the fields. We generically expect that a truncation of 8th will be sufficient to provide quantitatively reasonable estimates for the critical exponents.

Our present results clearly highlight the strength of a combination of the functional RG with the $\epsilon$-expansion: 
The latter allows us to compile a complete list of all FPs that can be continuously connected to the Gaussian FP at $d = 4$, whereas the former provides us with a quantitatively more reliable estimate of the stability regions of the different FPs. 
Combined, these methods allow us to arrive at a complete picture of stable FPs in the space of the $N_i$ while minimizing the computational effort.
%


\section{Conclusions}

With this study we identify a new fully-coupled FP in the $d = 3$ dimensional three- and four-field models. While we find that this FP is indeed IR-stable for some values of $N_i$, it does not lie at real fixed-point values for the couplings at other values of the $N_i$. This is in stark contrast to the $O(N)$ theory or the class of $O(N_1)\oplus O(N_2)$ models \cite{Fisher:1974zz,Kosterlitz:1976zza,Aharony:2002,Aharony:2002zz,Calabrese:2002bm,Folk:2008mi} where the relevant scaling solution(s) are either IR-stable or can be reached via additional fine-tuning. In addition to identifying a new FP, we confirm our previous finding \cite{PhysRevE.90.052129} that for certain multi-field models there is no IR-stable multicritical scaling solution. This behavior is directly tied to the properties of the new fully coupled fixed points, the ACFP and the BIFP.
Thus, this study has further clarified the reason for the absence of multicritical scaling solutions: While in the two-field model \cite{Fisher:1974zz,Kosterlitz:1976zza,Aharony:2002,Aharony:2002zz,Calabrese:2002bm,Folk:2008mi} different FPs exchange stability \emph{only} through a collision of two fixed points at real values of $N_i$, the three- and higher-field models feature the \textit{additional} possibility that FPs emerge from the complex plane. 

Our study plays out the strengths of two methods:
The $\epsilon$-expansion allows for a straightforward identification of all FPs that can be continuously connected to the Gaussian FP in $d = 4$ dimensions, as the $\beta$-functions are polynomial in the couplings.
In contrast, the fixed-point search is more involved with the functional RG due to the non-polynomial nature of the $\beta$-functions. 
However, the functional RG provides better quantitative results already at low orders of the LPA. This can be seen clearly for the example of the DFP. To estimate its stability regime, we may apply an \emph{exact} scaling relation \cite{PhysRevLett.51.2386, Aharony:1974, Aharony:1976, Aharony:2002, Aharony:2002zz} to determine the exponent $\theta_4$ from critical exponents of the $O(N)$ Wilson-Fisher FP \cite{LeGuillou:1977ju,Guida:1998bx,Wilson:1973jj,Wegner:1976bk,Pelissetto:2000ek}.
By doing so, we find that the result from the LPA at 4th order in the fields provides a quantitatively more reliable estimate for the scaling exponent than the $\epsilon$-expansion at one-loop order.
Taken together, the two methods thus allow for an efficient identification of all existing FPs, using the $\epsilon$-expansion at low orders, followed by a leading order determination of the stability regimes and critical exponents with the functional RG.

The identification of distinct interacting FPs in three- and four-field models also allows us to explore RG trajectories that define both UV- and IR-complete QFTs in $2 < d < 4$ dimensions. In general, multi-field theories provide a large number of such trajectories and typically feature two distinct regimes when it comes to the question of symmetry enhancement in the IR: For values of $N_i$ where a symmetry-enhanced FP is IR stable, all other FPs require a higher degree of fine-tuning to reach them in the IR. Thus, IR symmetry enhancement appears as a ``natural'' possibility that requires the least amount of fine tuning. In contrast, for other values of $N_i$, the same symmetry-enhanced FP will feature additional relevant directions, giving rise to the familiar notion that an enhancement of symmetry typically requires additional fine-tuning.

Our findings might have implications for possible UV completions of coupled scalar models in $d = 4$ dimensions. We observe that the asymmetrically coupled FP can be found in the $\epsilon$-expansion, \textit{i.e.}, it emerges from the Gaussian FP at $d < 4$.
Thus we conclude that no nontrivial FP exists for these models in $d=4$, unless it lies within a strongly nonperturbative regime at very large values of the couplings. 
This implies that, e.g., inflationary models with several scalar fields are not UV complete, but instead most probably feature Landau poles at finite scales.
Interestingly, a coupling to gravity could facilitate a UV completion in the context of asymptotically safe models. Studies suggest that a gravitational FP persists when the effects of several minimally coupled scalars are taken into account \cite{Dona:2014pla}. It is of course interesting to understand whether a similar statement applies to interacting matter models. 
In particular, the new universality classes that we discuss in this paper and which are inherent to $n$ field models ($n\geq 2$) could potentially survive an extension to 4 dimensions, when gravitational effects are added, as these generically seem to shift Gaussian FPs to interacting FPs \cite{Eichhorn:2012va}. Thus gravity might extend the upper critical dimension for this interacting FP to $d>4$.
Following the methods discussed in \cite{Percacci:2015wwa,Borchardt:2015rxa,Labus:2015ska}, an assessment of this scenario could be possible. In the context of scalar dark-matter models, where a coupling to other matter fields is less relevant, the existence of such scalar-gravity FPs could provide a predictive UV completion.
%


\vskip 10pt

{\it Acknowledgments} 
A.E. acknowledges support by the Perimeter Institute for Theoretical Physics through an Emmy Noether fellowship during the initial stages of this project.
The work of A.E. is supported by an Imperial College Junior Research Fellowship.
D.M. is supported by the European Research Council under the European Unions Seventh Framework Programme (FP7/2007-2013) / ERC grant agreement 339220.
M.M.S. is supported by the grant ERC-AdG-290623.


\begin{appendix}

\begin{table}[!b]
\setlength{\tabcolsep}{4.7pt}
\renewcommand{\arraystretch}{1.3}
\begin{tabular}{ccc|cccccc}
$N_1$ & $N_2$ & $N_3$ & $\lambda_{200}$& $\lambda_{020}$& $\lambda_{002}$& $\lambda_{110}$& $\lambda_{101}$& $\lambda_{011}$\\ \hline \hline
1 & 1 & 1 & 0.084 & 0.084 & 0.068 & 0.084 & 0.115 & 0.115 \\ \hline\hline
1 & 2 & 2 & 0.085 & 0.085 & 0.091 & 0.085 & 0.054 & 0.054 \\ \hline \hline
\end{tabular}\\
\vspace{0.1cm}\setlength{\tabcolsep}{4.7pt}
\begin{tabular}{ccc|cccccc}
$N_1$ & $N_2$ & $N_3$ & $\theta_4$ & $\theta_5$ & $\theta_6$ & $\theta_7$ & $\theta_8$ & $\theta_9$\\ \hline \hline
1 & 1 & 1 & 0.130 & -0.011 & -0.101 & -0.491 & -0.708 & -1 \\ \hline\hline
1 & 2 & 2 & 0.097 & -0.017 & -0.053 & -0.667 & -0.836 & -1 \\ \hline\hline
\end{tabular}
\caption{\label{tab3}Fixed-point values and critical exponents for the symmetry-enhanced BIFP. A positive exponent $\theta_4$ indicates that this FP is not IR-stable.}
\vskip 22pt
  \setlength{\tabcolsep}{12pt}
  \renewcommand{\arraystretch}{1.3}
\begin{tabular}{cc|ccc}
$N_1$ & $N_2$  & $\lambda_{20}$& $\lambda_{02}$&  $\lambda_{11}$\\ \hline \hline
1 & 2 & 0.084 & 0.067 & 0.115 \\\hline\hline
3& 2& 0.085 & 0.091 & 0.054 \\\hline\hline 
\end{tabular}\\
\vspace{0.1cm}
\setlength{\tabcolsep}{12.5pt}
\begin{tabular}{cc|ccc}
$N_1$ & $N_2$ & $\theta_3$ & $\theta_4$ & $\theta_5$ \\ \hline \hline
1 & 2 & 0.130 & -0.491 & -1 \\\hline\hline
3 & 2 & -0.053 & -0.836 & -1 \\ \hline\hline
\end{tabular}
\caption{\label{tab4}For comparison, we show fixed-point values and critical exponents for the biconical FP in the two-field case. There are only \emph{two} relevant directions in this model and it is therefore the exponent $\theta_3$ that decides about the IR-stability of the scaling solution.}
\end{table}

\section{Fully-coupled FPs without IR-stability}
\label{Sec:Appendix}

Here, we discuss the existence of FPs, that feature symmetry-enhancement to $O(N_i+N_j)\oplus O(N_k)$ symmetry.
Unlike the DIFP, these FPs feature nonvanishing interactions that couple the $(i,j)$- to the $k$-sector, which does not take part in the symmetry-enhancement. 
We can identify these as arising from the biconical FP in the two-field model, \textit{i.e.}, their coordinates are given by the coordinates of the two-field BFP for $N_1+N_2$ and $N_3$ (and the two other combinations, respectively).
As an example, let us focus on the $O(N_1+N_2)\oplus O(N_3)$-symmetric FP.
To reach this FP, it is necessary to tune at least one additional direction. 
As an example, we give the coordinates and critical exponents of the FP at several selected points, cf.\ Tab.\ \ref{tab3}.
Comparing the critical exponents to those of the BFP in the two-field model we realize that three of the critical exponents are inherited from the BFP. 
For comparison, we list the fixed-point coordinates and critical exponents for the two-field BFP at the corresponding points, cf.\ Tab.\ \ref{tab4}.

\begin{figure}[!t]
\includegraphics[width=0.49\linewidth]{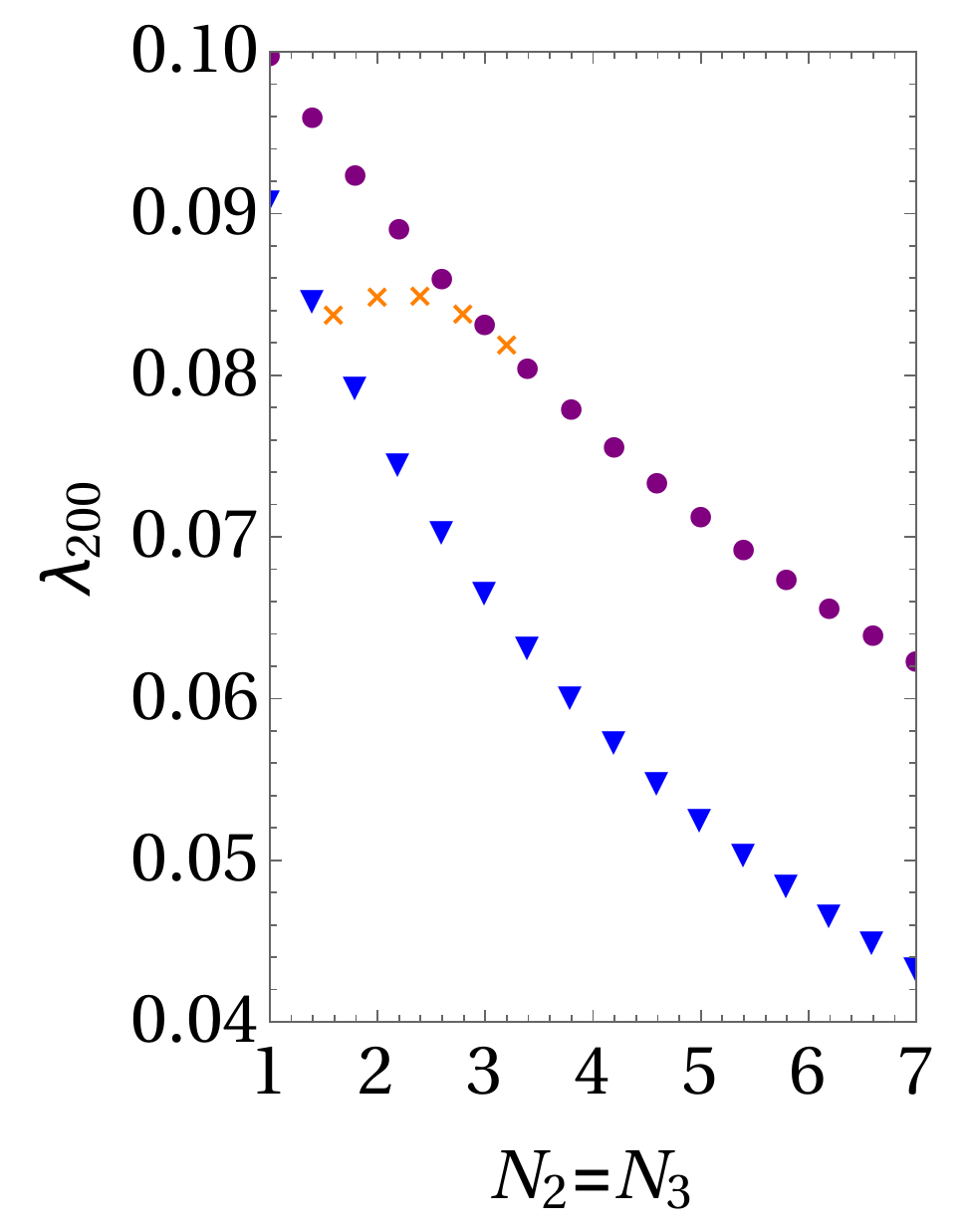}\includegraphics[width=0.49\linewidth]{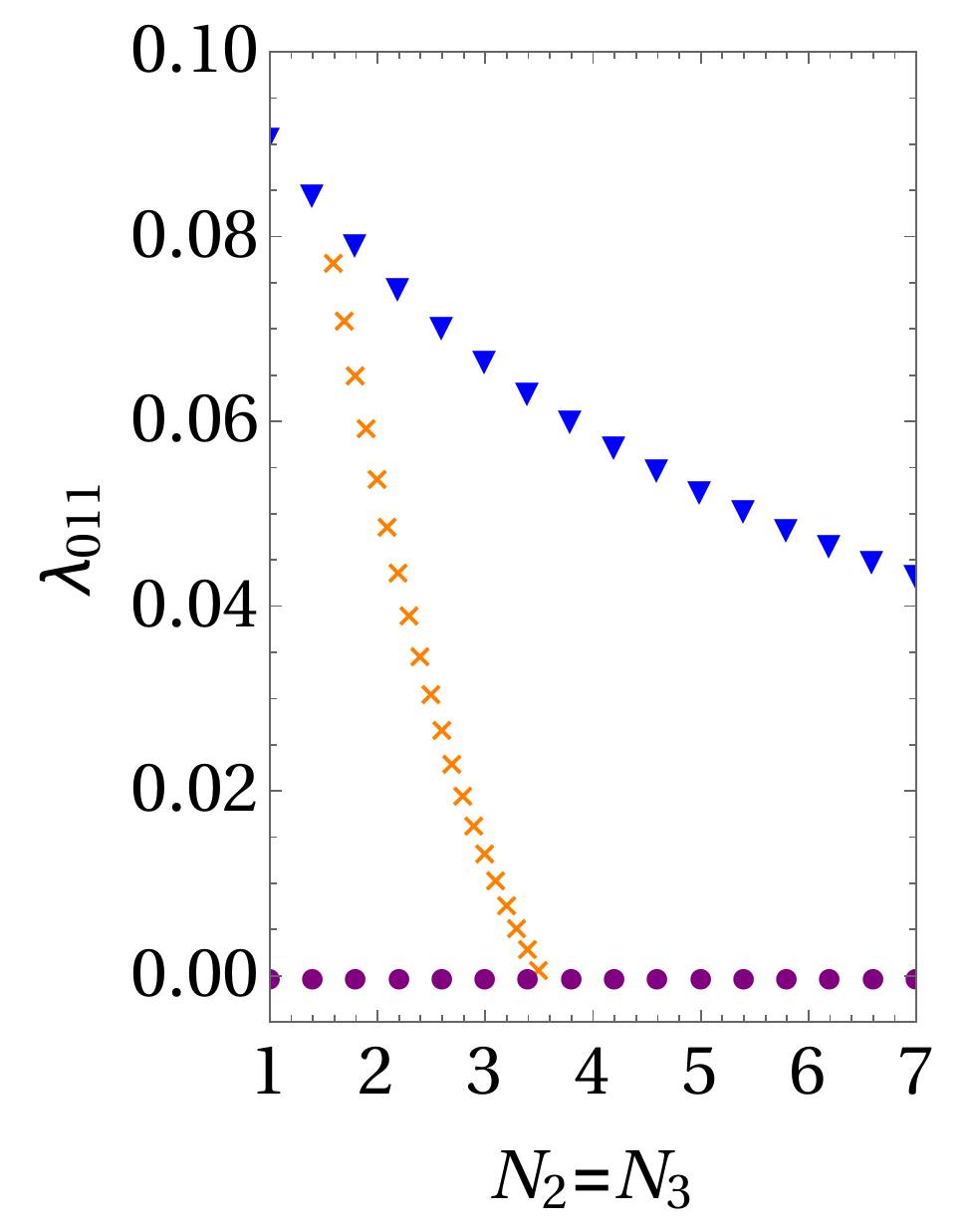}
\caption{\label{fig4}We show the IFP ({\small{\color{blue}$\blacktriangledown$}}), the DIFP ({\color{purple}$\bullet$}), and the coupled isotropic BIFP ({\color{orange}\small$\times$}) as a function of $N_2 = N_3$ for fixed $N_1 = 1$. The BIFP collides with the IFP at $N_2 = N_3 = 1.5$ and renders the IFP unstable without becoming stable itself. At larger values of $N_2 = N_3$ it collides with the DIFP. The coupled BIFP is shown only between the collision points to illustrate its role as a mediator between the IFP and DIFP.}
\end{figure}
\begin{figure}[!t]
\includegraphics[width=0.7\linewidth]{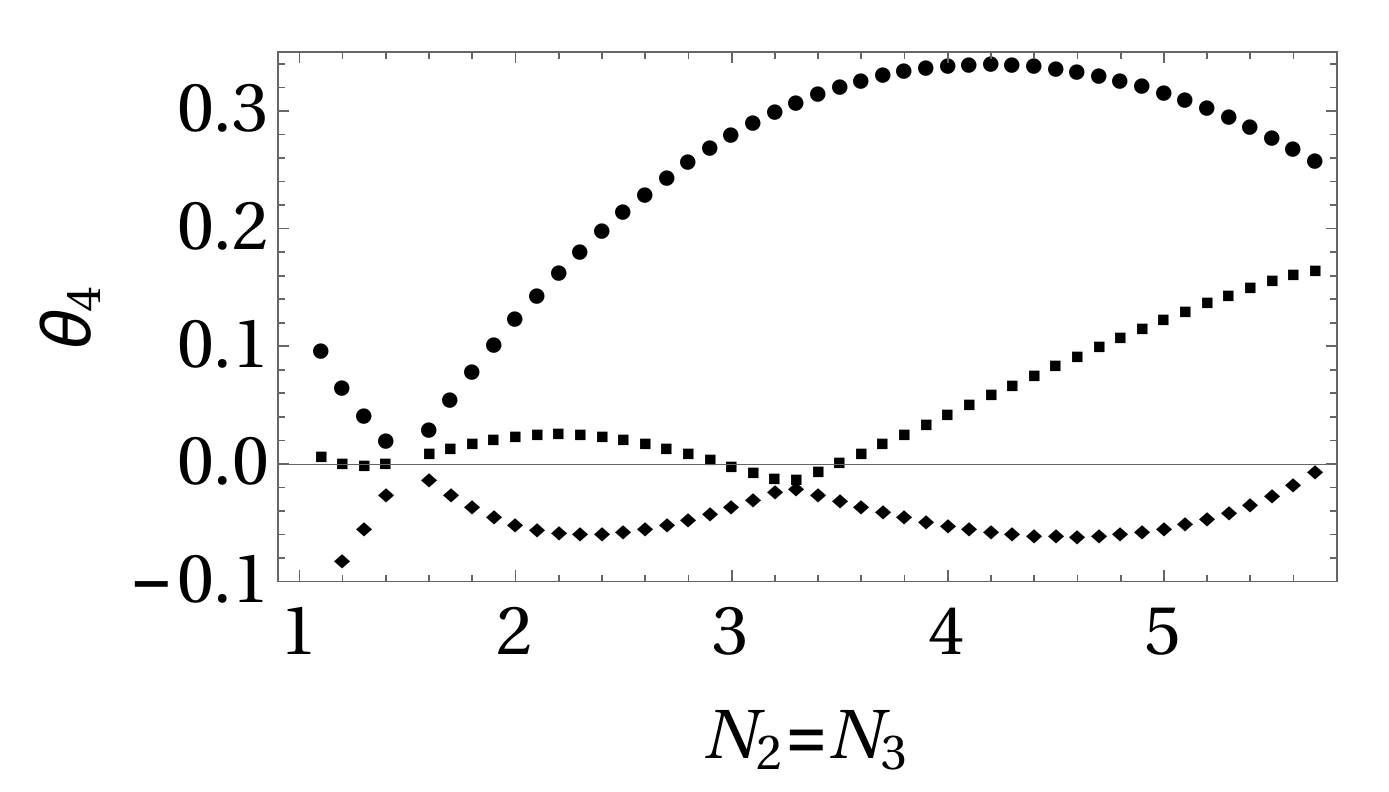}\\[-5pt]
\hspace{-0.2cm}\includegraphics[width=0.72\linewidth]{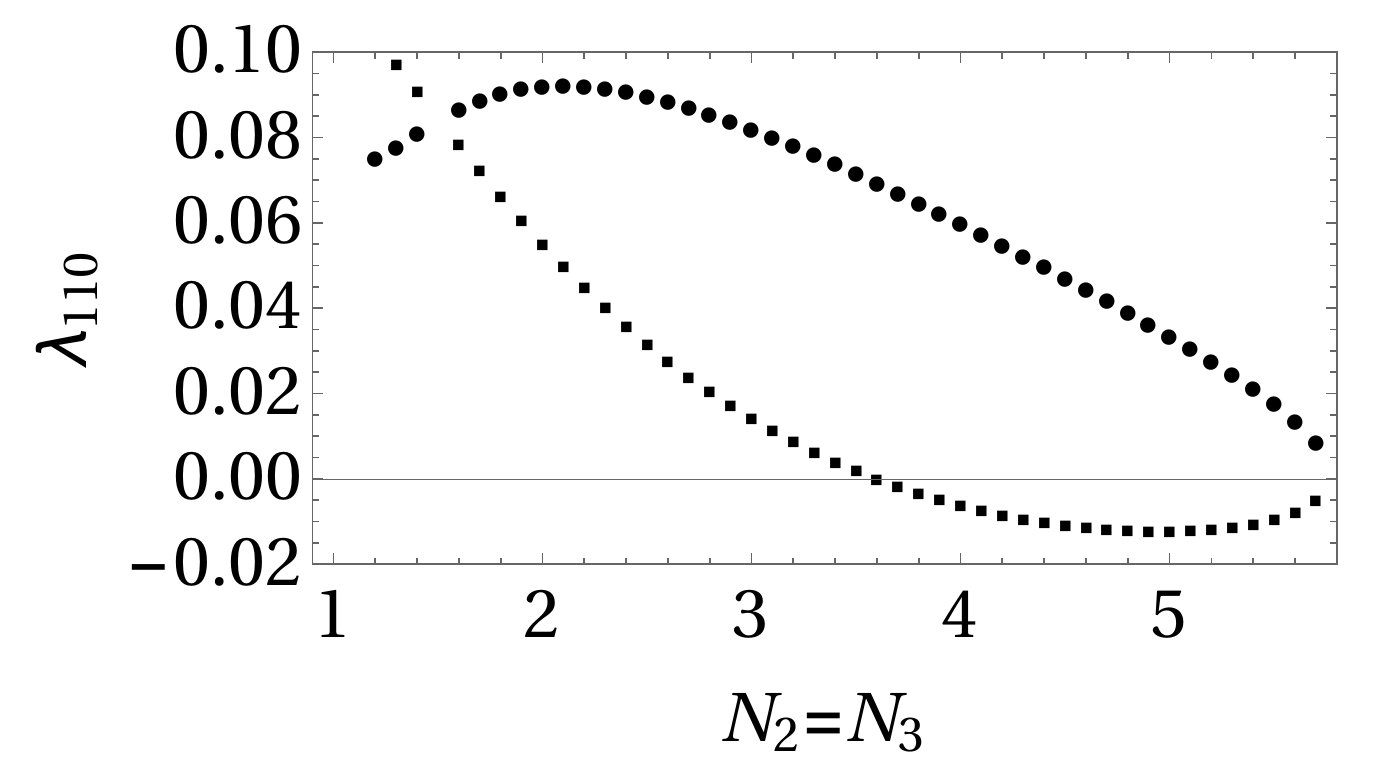}
\caption{\label{fig5}Two new, fully coupled FPs also collide with the IFP at $N \equiv N_2 = N_3 = 1.5$. These two FPs are mapped into each other under the exchange of $\phi_2$ and $\phi_3$. Thus their critical exponents are identical, as are some of their couplings. We show the the values of $\lambda_{110}$ for the two FPs (lower panel), to illustrate their collision at $N\approx 5.8$, where they disappear into the complex plane.}
\end{figure}

Two of these FPs collide with the isotropic FP at $N_1=1$, $N \equiv N_2=N_3=1.5$ and destabilize the IFP. 
Additionally, two more coupled, non-symmetric FPs are involved in this collision. 
These new FPs are nowhere stable. 
They disappear into the complex plane away from the real axis at $N \approx 5.8$, where they collide with each other. 
We show their critical exponents and selected fixed-point coordinates in Fig.\ \ref{fig5}.
The collision that destabilizes the IFP leaves no stable FP. 
The coupled IFP then moves on to collide with the DIFP which features a similar symmetry-enhancement to an $O(N_1 + N_2) \oplus O(N_3)$ symmetry, cf.\ Fig.\ \ref{fig4}. 
The existence of new, symmetry-enhanced FPs is thus responsible for the early destabilization of the IFP.

\end{appendix}

\newpage

\bibliography{references}

\end{document}